\documentclass[conference]{IEEEtran}
\IEEEoverridecommandlockouts
\usepackage{cite}
\usepackage{amsmath,amssymb,amsfonts}
\usepackage{algorithmic}
\usepackage{graphicx}
\usepackage{textcomp}
\usepackage{xcolor}
\usepackage{glossaries}
\usepackage[hyphens]{url}
\usepackage[hidelinks]{hyperref}

\def\BibTeX{{\rm B\kern-.05em{\sc i\kern-.025em b}\kern-.08em
    T\kern-.1667em\lower.7ex\hbox{E}\kern-.125emX}}

\newacronym{anb}{AnB}{Alice \& Bob}
\newacronym{uml}{UML}{Unified Modeling Language} 
\newacronym{mde}{MDE}{Model-Driven Engineering} 
\newacronym[plural = DLTs, firstplural=Distributed Ledger Technologies (DLTs)]{dlt}{DLT}{Distributed Ledger Technology} 
\newacronym{ics}{ICS}{Industrial Control System}
\newacronym{qos}{QoS}{Quality of Service}
\newacronym{iot}{IoT}{Internet of the Things}
\newacronym{bnf}{BNF}{Backus-Naur Form}
\newacronym{ertms}{ERTMS}{European Railway Traffic Management System}
\newacronym{etcs}{ETCS}{European Train Control System}
\newacronym{kms}{KMS}{Key Management System}

\begin{document}

\title{Model-Driven Engineering for Formal Verification and Security Testing of Authentication Protocols
}

\author{\IEEEauthorblockN{Mariapia Raimondo}
\IEEEauthorblockA{\textit{Dip. di Matematica e Fisica} \\
\textit{Universit\`a della Campania  ``L. Vanvitelli''}\\
Caserta, Italy \\
mariapia.raimondo@unicampania.it}
\and
\IEEEauthorblockN{Stefano Marrone}
\IEEEauthorblockA{\textit{Dip. di Matematica e Fisica} \\
\textit{Universit\`a della Campania  ``L. Vanvitelli''}\\
Caserta, Italy \\
stefano.marrone@unicampania.it}
\and
\IEEEauthorblockN{Angelo Palladino}
\IEEEauthorblockA{\textit{Aerospace Business Unit} \\
\textit{Kineton srl}\\
Napoli, Italy\\
angelo.palladino@kineton.it}

}

\maketitle

\begin{abstract}
Even if the verification of authentication protocols can be achieved by means of formal analysis, the modelling of such an activity is an error-prone task due to the lack of automated and integrated processes. This paper proposes a comprehensive approach, based on the \gls{uml} profiling technique and on model-transformation, to enable automatic analysis of authentication protocols starting from high-level models. In particular, a UML-based approach is able to generate an annotated model of communication protocols from which formal notations (e.g., AnBx, Tamarin) can be generated. Such models in lower-level languages
can be analysed with existing solvers and/or with traditional testing techniques by means of test case generation approaches. The industrial impact of the research is high due to the growing need of security and the necessity to connect industrial processes and equipment to virtualised computing infrastructures. The research is conducted on two case studies: railway signalling systems and blockchain based applications. 
\end{abstract}

\begin{IEEEkeywords}
Formal verification of security protocols, Model Driven Engineering, Verification and Validation, Railway signalling, blockchain technology
\end{IEEEkeywords}

\section{Introduction}\label{sec:intro}

Critical systems are now connected to the Internet due to an increase in the complexity of their functionalities. This growth of complexity goes not only in the direction of increasing the level of automation but also in opening such systems to remote control. 
Security becomes a prime factor in such a context: its lack could lead not only to economic loss or privacy leaks, but also to damage to people and goods. This is the case of the railway signalling systems, which is considered in this paper.

Another emerging domain where security plays an important role is the one constituted by the emerging \glspl*{dlt} and, despite its relatively recent adoption, a lack of security in \gls*{dlt} would have a strong social impact. In fact, the blockchain technology, which is a particular case of \gls{dlt}, has received widespread support and acclamation.
It provides an infrastructure to manage transactions within a community without the need of a trusted third party that supervises. Suffice to say, cryptocurrencies are based on blockchain technology and they are one of the main strategies of long-term investment adopted nowadays. Moreover, the European Community has started writing down regulations\footnote{\url{https://eur-lex.europa.eu/legal-content/EN/TXT/?uri=CELEX\%3A32022R0858&qid=1656931726550}} to use them and proposals to prevent the usage of this technology for illicit purposes\footnote{\url{https://eur-lex.europa.eu/legal-content/EN/TXT/?uri=CELEX\%3A52021PC0422&qid=1656931726550}}.

Formal verification can be very useful both to check the correctness of the behaviour of \glspl{ics} and the achievement of the security levels required by the new technologies such as the blockchain ones. Unlike simulation and testing, this complementary technique is able to find very specific conditions that could bring to a security flaw that has not been considered in security test plans. Furthermore, formal methods are recommended for certification purposes, especially in critical systems.
Nonetheless, it is worth underlining that: simulation and testing are still the most effective methods to demonstrate the presence of a security issue in a specific scenario; formal modelling and analysis often require specialised skilled people, whose effort is devoted to low-level details error-prone activities.
The work presented in this paper deals with the problem of easing the work of the modeller and unifying the approach to check security and behavioural properties. The main objective is to provide a comprehensive approach, supporting formal analysis and testing, based on \gls*{mde} techniques.

The approach presented in the paper is based on a traditional model-driven process schema with the following elements. First, a \gls*{uml} Profile able to capture the authentication related features of the modelled system is defined. Then, a model transformation is provided to derive a formal notation from an annotated \gls*{uml} model of the system. Finally, the produced model is analysed with a set of techniques to verify security properties by formal analysis and/or to generate test scripts to support the verification. Where possible, the approach involves the use of existing and assessed solvers and toolchains.
More in particular, the \gls*{uml} Profile is used to enrich behavioural models with cryptoprimitives and security properties. The model transformation is used to derive an \gls*{anb} model which is then checked over security properties. At the best of our knowledge, there is no unifying description framework for the different tools that could be used in the formal verification approach. This work will be
also driven
by two case studies: the \gls*{ertms}/\gls*{etcs} \gls*{kms} and the Tweetchain protocol.

The paper is structured as follows: Section~\ref{sec:rel} gives a quick revision of related works, Section~\ref{sec:over} describes the proposed approach, while Section~\ref{sec:tow} discusses about specific technical concerns. Finally, Section~\ref{sec:prog} resumes the current developmental state of the approach.

\section{Related works}\label{sec:rel}
As mentioned in Section~\ref{sec:intro}, one of the adopted methodologies is the \gls*{uml} profiling technique. In particular, up to now, there is no \gls*{uml} standard profile --- i.e., a \gls*{uml} profile defined by the Object Management Group --- devoted to the security analysis of blockchain-based protocols and applications, which we aim to propose. However, regarding \gls*{uml} profiles for security analysis, the QoS\&FT profile~\cite{QoSFT} provides general support for the specification of \gls*{qos} characteristics and for risk assessment. Moreover, many researchers contributed by proposing \gls*{uml} profiles useful for the modelling and analysis of security properties of software systems, such as 
\gls*{uml}Sec~\cite{Jurjens05} and SecAM~\cite{SecAM}.

From a model-driven perspective, the scientific community has traditionally proposed many model-based approaches that are specific to one simulation platform per time (e.g., \cite{Anastasakis2007436}). Moreover, a set of contributions regarding the test case generation from high level models is present (e.g., \cite{Utting2007}).
The combination of both formal modelling and test case generation is present in a series of works that use, as an example, the model checking technique, as in \cite{10.1145/318774.318939}.

Regarding the railway domain, security is rarely approached in scientific studies, which more frequently focus on other non-functional features as safety, availability, performance and signalling. In~\cite{Bougacha2019367} the authors provide an approach to model and analyse the railway signalling system in Event-B. Similarly, in~\cite{Marrone2014669} a process is presented to generate Generalized Stochastic Petri Net (GSPN) and Promela models for analysis and test case generation purposes. In~\cite{Scippacercola201731}, instead, a model-based approach is provided that is aimed at improving the development of proprietary railway interlocking systems. 

On the other side, security analysis in blockchain environments by means of formal methods mainly focuses on smart contracts. The survey~\cite{SinghPZCD20} discusses 35 papers from 2015 to 2019 just focusing on formalization of smart-contracts.

\section{Overview of the approach}\label{sec:over}
\begin{figure}[htbp]
\centerline{\includegraphics[width=0.95\columnwidth]{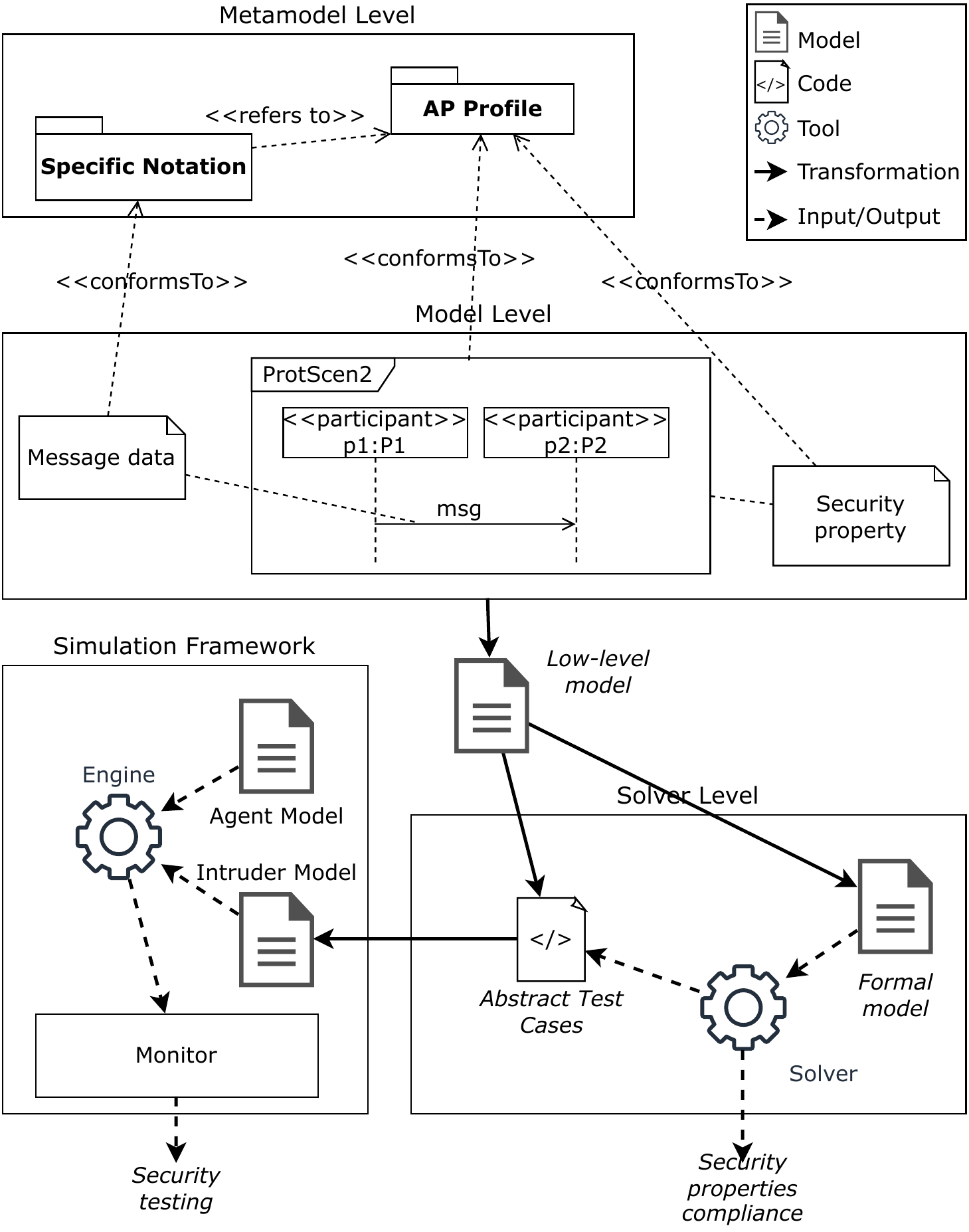}}
\caption{Overview of the approach}
\label{fig:approach}
\end{figure}

Figure~\ref{fig:approach} describes the approach presented in the paper. The aim of the approach is to facilitate the verification of security properties of communication protocols. This goal is achieved by allowing the definition of the protocol, and of the related security properties to be verified, in a high-level language (\textit{Model level}). 
Moreover, we leverage model-driven techniques to 
automatically generate low-level artefacts and run
verification activities. The latter can be carried out
using formal and simulation techniques (\textit{Solver level} and \textit{Simulation Framework}).
A key feature of this approach is the possibility of using different platforms to analyse protocols, that range from formal verification tools to test scripts generation ones. 

The chosen high-level modelling language is the well-known \gls*{uml}, which provides a set of different diagrams that can be used to model a given system. In particular, dealing with communication protocols, good options are the Sequence Diagrams and the protocol State Machine Diagrams. The first capture the interactions between parties, whilst the second focus more on the state changes of participants in reaction to message exchanges.

Since our aim is to ease the work of the modeller, a way to fill the gap that arises from the creation of a \gls*{uml} model of a protocol is to provide more protocol-related concepts at \gls*{uml} level.
This can be done by applying the \gls*{uml} Profiling technique to capture and define as a Profile the authentication related features of the modeled systems.

In Fig.~\ref{fig:approach}, \gls*{uml} and the \textit{AP Profile} --- Authentication Protocol Profile --- constitute the high-level language that can be used to model both the protocol specification and the security properties to be verified (e.g., by means of \gls*{uml} constraints or other annotated \gls*{uml} constructs).
Then, thanks to a model transformation, both the protocol and the security properties are translated in a \textit{Low-level model}. 
The latter is a tool-independent notation, for example \gls*{anb}. 
Finally, further model transformations are used to obtain formal models and a set of specific solvers can be used:
\begin{enumerate}
    \item To formally check the model against the translated security properties;
    \item To get counterexamples of the unmatched properties to extract \textit{Abstract Test Cases (ATCs)} from such counterexamples;
    \item To transform, according to the presence of a 
    system simulator framework, the \textit{ATCs} into an intruder model;
    \item To simulate such model (i.e., intruder and protocol participants) to get a practical demonstration of the presence of the security flaw.  
\end{enumerate}

\section{Towards an implementation of the approach}\label{sec:tow}

This section discusses three technical concerns to match in order to realize the approach. 

\subsection{Reusabilty}
One of the key goals of the proposed research is to define a methodology that can be easily adapted in different domains.
The sequence of the exchanged messages, and the type of information they carry, can differ a lot among the various authentication protocols.
This difference is already present in classical protocol descriptions (i.e., standards, white papers, technical reports, etc.). The separation of these descriptions boosts the flexibility of the approach. 
The concrete solution in this work considers two sub-languages.
 
The first is represented by a classical \gls*{uml} Profile: an example of the concepts represented at this level are the message, that can be captured in the \gls*{uml} Profile by a stereotype to annotate Sequence Diagram's messages. At the second level, there are instead specific textual notations, formalised by \gls*{bnf} grammars. These notations can be used to detail which are the data exchanged (both in plain and cypher-texts) between communication parties.

The \gls*{uml} Profile is separated from the specific notation and, thus, it can be reused in different specific domains. When a modeller specifies a protocol, he/she can use the stereotypes considered in the \gls*{uml} Profile to model and annotate the \gls*{uml} diagrams accordingly. Moreover, as tagged values of \gls*{uml} stereotypes give the possibility to add textual information, they are used with the \textit{Transaction} stereotype. By doing so, the text can be formatted according to the specific grammar defined for the specific domain. A preliminary design of this mechanism is published in~\cite{ijchvt}.

\subsection{Choice of the solution toolchain}
One of the most valuable benefits coming from the adoption of \gls*{mde} is that it firmly separates the level of abstraction of the different artefacts. 

Part of the implementation of the toolchain presented in this paper can be clearly achieved reusing existing tools and integrating part of existing project in ours. 
In particular, the work of Modesti~\cite{BUGLIESI201646} caught our attention, since it provides the generation of analysable formal models starting from semi-formal language models. 
The authors in a first phase of their work developed a on-the-fly model checker called OFMC (Open-Source Fixed-Point Model-Checker). It is a model checker naturally devoted to the analysis of security protocols and it requires as input a model written in AnB/AnBx language. This makes it an adequate component to be used as partial back-end of our proposal.

In detail, the AnBx language is a formalized version of the well-known semi-formal \gls*{anb} notation. It provides a higher level of abstraction by providing a easier way to model cryptographic primitives in the shape of a \textit{channel}. In fact, the \textit{mode} of a \textit{channel} allows the modeller to enforce or weaken the security guarantees of messages exchange in a convenient style, decoupling the protocol from the specific encryption method adopted.

It is valuable to underline that the Modesti's toolchain does not represent the only possible choice. In~\cite{forse} and in~\cite{ijchvt}, the Tamarin Prover model checker has been used. In particular, in~\cite{forse}, Tamarin has been used as a single language (i.e., detached by a \gls*{mde} approach), while in~\cite{ijchvt} a model-driven toolchain is proposed using a generic \gls*{anb} notation and Tamarin as a solving back-end.

\subsection{Integrating formal analysis and testing}
The community of security testing is wide and it considers finding vulnerabilities and errors in low-level code and operating systems as well as finding high-level protocol flaws. The formal methods and model-driven bodies of knowledge are more prone to support the second kind of activity and, thus, talking about the possibility of implementing security testing features in our approach could seem strange. 

As Fig.~\ref{fig:approach} reports, the testing ``branch'' of the approach (from \textit{Abstract Test Cases}) relies on the presence of a \textit{Simulation Framework} whose purpose is to define an executable abstraction of the behaviours of protocol parties. At this point of the research, there is no specific requirement for such frameworks: they could span from simple event based simulation to complex hardware/system-in-the-loop environments. 

A possible architecture for the \textit{Simulation Framework} is here discussed.
The core of the architecture is a simulation engine executing the participant models, which mainly exchange messages, and the intruder, which attempts to intercepts such messages injecting malicious ones. The engine is supported by these components: (1) a translator from the \textit{ATCs} and security property to the concrete notation for the test script; (2) a security monitor, which is in charge of monitoring the security properties, raising exceptions in case of property violations. The \textit{Simulation Framework} is based on the presence of a language for test scripting to express the models of the actors and of the security properties.

\section{State of progress}\label{sec:prog}
We plan to prototype and apply the approach here presented to two different case studies: one is the Tweetchain protocol, the other one is the Key Management System of the ERTMS/ETCS.

The Tweetchain protocol, described in~\cite{Buccafurri2017}, is a lightweight version of the blockchain paradigm. It mainly consists in a reinvention of the consensus protocol to make the protocol applicable to \gls*{iot} devices, which do not own the necessary computing power and memory to perform one of the full consensus protocols available. In fact, the authors leveraged the feature of \gls*{iot} devices of being connected to the Internet to base the consensus protocol on the famous social network Twitter, by using tweets, to encode transactions, and meshed replications, to substitute the proof of work. 
Currently, foundations were laid for the creation of a blockchain-specific \gls*{uml} Profile, which could be a first version of the \textit{AP Profile}. Moreover, a specific textual notation for the Tweetchain data exchange has been provided, that is expressed by a \gls*{bnf} grammar.

Furthermore, a transformation from a UML Sequence Diagram annotated with such a profile to the AnB notation is designed. Finally, thanks to the already available automatic translation provided by~\cite{keller}, it has been possible to translate the \gls{anb} protocol into a Tamarin model~\cite{Meier2013696}. Only formal analysis is conducted up to now on this protocol. Previous published papers describing the preliminary results on this case study are~\cite{forse,ijchvt}.

To guarantee the security of signalling systems in modern railway, Key Management System regulates the exchanges of the keys between trackside and onboard controllers. The ETCS-KMS protocol is devoted to guarantee the security of this exchange in an on-line environment~\cite{subset137:std}.

In~\cite{tesimariapia}, a preliminary study on the application of the techniques here presented is reported, focusing on a compositional approach of Tamarin specification.

\begin{figure}[htbp]
\centerline{\includegraphics[width=0.9\columnwidth]{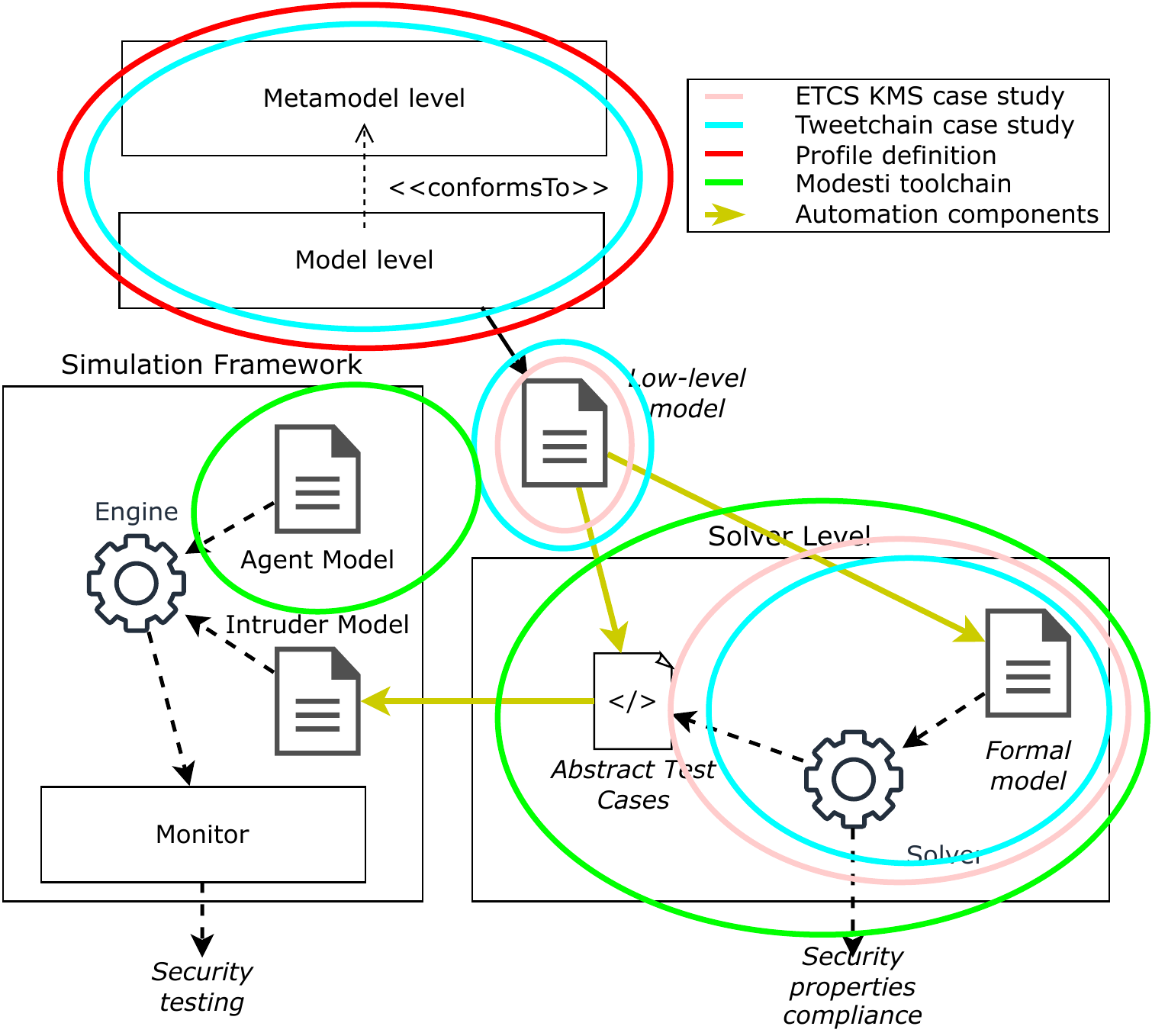}}
\caption{Roadmap of the work}
\label{fig:final}
\end{figure}

Fig.~\ref{fig:final} wants to summarise the status of the development of the entire approach. It recalls Fig. \ref{fig:approach}. The blue circles represent the part of the research covered by the works about the Tweetchain protocol \cite{forse,ijchvt}, while the pink ones are related to the ETCS-KMS case study \cite{tesimariapia}.

The next steps in this research will be: (1) the migration of the case studies to the toolchain based on the work of Modesti, to exploit the translator from AnBx to OFMC and related tools (in green in the figure); (2) to define the blockchain-specific Profile and extend it to the \textit{AP Profile}, to include also the ETCS-KMS specific features (in red); (3) the automation of the approach by means of the prototyping of key components (in yellow).

In particular, the original contributions of this work will be: (1) the definition of a \gls{uml} Profile for blockchain-based applications and authentication protocols; (2) the integration of an intruder model in the \textit{Simulation Framework}, which is not present in any of the considered tools; (3) the provision of a framework giving the evidence of attacks arisen from formal analysis.

\section*{Acknowledgment}

The work of Mariapia Raimondo is granted by INPS --- Istituto Nazionale di Previdenza Sociale (Italy) --- with the PhD program XXXVI cycle.

\bibliographystyle{IEEEtran}
\bibliography{biblio.bib}

\end{document}